\documentclass[preprint,showpacs,preprintnumbers,amssymb,aps,nofootinbib]{revtex4}
\usepackage{amsfonts,color,graphicx,epsfig,graphicx,amsmath,multirow,slashed}

\begin{document}

\title{Next-to-leading order analysis of $J/\psi + \gamma$ production in photon-photon collisions at CEPC}

\author{Ying-Zhao Jiang}
\email{yjjiang@163.com}
\author{Zhan Sun}
\email{sunzhan_hep@163.com}

\affiliation{
\footnotesize
School of Physics and Mechatronics Engineering, Guizhou Minzu University, Guiyang 550025, People's Republic of China. \\
}

\date{\today}

\begin{abstract}

We systematically investigate the production of $J/\psi + \gamma$ in $\gamma\gamma$ collisions within nonrelativistic QCD (NRQCD) factorization, with the direct-photon channel calculated specifically up to the next-to-leading order in $\alpha_s$. Calculations for CEPC energy region show the resolved photon contribution is negligible, while the direct photon process dominates, yielding substantial annual $J/\psi$ yields. Significant modifications to $J/\psi$ polarization parameters emerge from color-octet mechanisms, and different NRQCD long distance matrix elements (LDMEs) further yield distinct polarization patterns. Furthermore, the polarization predictions are highly sensitive to the $^3P_J^{[8]}$ LDME, while being insensitive to the $^1S_0^{[8]}$ and $^3S_1^{[8]}$ LDMEs. Leveraging the cleaner environment of $e^+e^-$ collisions versus hadronic processes, the production of $J/\psi$ associated with a photon in $\gamma\gamma$ collisions provides a high-precision platform to test LDMEs universality and resolve longstanding $J/\psi$ polarization puzzles.
\pacs{13.66.Bc, 12.38.Bx, 12.39.Jh, 14.40.Pq}

\end{abstract}

\maketitle

\section{Introduction}{\label{intro}}

Non-relativistic QCD (NRQCD) factorization \cite{Bodwin:1994jh,Petrelli:1997ge} stands as one of the most effective frameworks for describing heavy quarkonium processes. It provides a systematic approach to disentangle the effects arising from dynamics at different scales. Specifically, the production process can be factorized into a sum of products involving short-distance coefficients (SDCs) and long-distance matrix elements (LDMEs). The SDCs are perturbatively calculable as a power series in $\alpha_s$, while the LDMEs are typically extracted by fitting theoretical predictions to experimental data. NRQCD has achieved considerable successes in describing numerous processes, particularly hadroproduction. For instance, NRQCD predictions bridged the large gap between leading-order (LO) color singlet model (CSM) predictions and the measurements of $J/\psi$ and $\psi(2S)$ hadroproduction reported by the CDF collaboration \cite{Braaten:1994vv,Cho:1995vh,Cho:1995ce}. Furthermore, measurements of $\chi_c$ and $\eta_c$ hadroproduction are also consistently described within NRQCD \cite{Ma:2010vd,Zhang:2014coi,Butenschoen:2014dra,Han:2014jya,Zhang:2014ybe}, respectively. Recent studies on $J/\psi$ electroproductions in deeply inelastic $ep$ scattering at HERA further indicate that CS contributions alone are insufficient to account for the measurements; incorporating color octet (CO) partonic processes significantly improves the agreement between theory and experiment \cite{Sun:2017nly,Sun:2017wxk,Lansberg:2019adr,Boer:2024ylx}.

Despite its successes in many areas, the NRQCD factorization faces significant challenges. One example is that NRQCD predictions for $e^+e^- \to J/\psi+X_{\textrm{non}-c\bar{c}}$ severely overshoot experimental measurements \cite{Zhang:2009ym}. Furthermore, the long-standing $J/\psi$ polarization puzzle remains unresolved within the NRQCD framework. For instance, CMS data in the rapidity region $0.6 < |y| < 1.2$ report a $J/\psi$ polarization of approximately 0.15 \cite{CMS:2013gbz}, while NRQCD predictions yield a value near 0.3 \cite{Han:2014jya}. Such a significant discrepancy challenges the consistency of an effective theory. This tension is further highlighted by the contrasting polarization predictions from different LDME sets, even though they all describe the production rates satisfactorily \cite{Brambilla:2010cs,Butenschoen:2012px,Chao:2012iv,Gong:2012ug,Shao:2012fs,Shao:2014fca,Bodwin:2014gia,Sun:2015pia}. Consequently, resolving the $J/\psi$ polarization remains one of the most challenging problems in high-energy physics, necessitating further investigation through diverse production channels.

In this article, we propose utilizing the process of $J/\psi$ production associated with an isolated photon via two-photon collisions to further investigate the polarization of $J/\psi$. At $e^{+}e^{-}$ colliders, such collisions are mediated by nearly on-shell virtual photons from incoming leptons, offering a cleaner environment than hadroproduction due to suppressed strong-interaction backgrounds. For the specific process of $\gamma\gamma \to J/\psi+\gamma$, the direct channel dominates over single- and double-resolved contributions. This suppression stems from CO LDMEs and the Parton Distribution Functions (PDFs) of gluon/quark in a resolved photon. In contrast, inclusive $J/\psi$ production in $\gamma\gamma$ collisions is dominated by single-resolved processes, introducing theoretical uncertainties from the poorly constrained gluon/quark PDFs in photon. As will be demonstrated, the $J/\psi$ polarization prediction reveals relevance of the $^3P_J^{[8]}$ state with negligible $^1S_0^{[8]}$ contributions, while $^3S_1^{[8]}$ is strongly suppressed by $\mathcal O (\alpha_s)$ higher-order effects. These features establish the $J/\psi+\gamma$ production in photon-photon collision as a $\textit{clean}$ probe for investigating $J/\psi$ polarization. Considering its designed high luminosity and collision energy at CEPC, which offers large potential for the production of a statistically significant sample of $J/\psi + \gamma$ events via photon-photon collisions, this study will focus on the CEPC \cite{CEPCStudyGroup:2018rmc,CEPCStudyGroup:2018ghi} as the proposed platform.

The process of $J/\psi$ production accompanied by a photon has been extensively studied across various collider environments \cite{Drees:1991ig,Doncheski:1993dj,Kim:1994bm,Roy:1994vb,Mirkes:1994jr,Kim:1996bb,Mehen:1996vx,Cacciari:1996zu,Ma:1997bi,Japaridze:1998ss,Mathews:1999ye,Kniehl:2002wd,Klasen:2004az,Kniehl:2006qq,Li:2008ym,Lansberg:2009db,Li:2014ava,Alimov:2024pqt,Keung:1983ac,Bodwin:2013gca,Bodwin:2014bpa,Huang:2014cxa,Bodwin:2016edd,Bodwin:2017wdu,Bodwin:2017pzj,Li:2009ki,Sang:2009jc,Braguta:2010mf,Patel:2011zty,Fan:2012vw,Fan:2012dy,Chao:2013cca,Chen:2013mjb,Li:2013nna,Chen:2013itc,Wang:2013ywc,Sun:2014kva,Xu:2014zra,Chen:2017pyi,Brambilla:2017kgw,Chung:2019ota,Li:2019ncs,Brambilla:2020xod,Yu:2020tri,Sang:2020fql,Liao:2021ifc,Wang:2023ssg}. Researches at hadronic colliders demonstrate significant discrepancies between NRQCD and CS predictions for the $pp \to J/\psi + \gamma + X$ process \cite{Kim:1996bb,Cacciari:1996zu,Li:2014ava}, particularly in polarization distributions \cite{Li:2014ava}. Studies at electron-proton colliders (e.g., HERA) have also revealed the crucial role of the CO mechanism in the $ep \to J/\psi + \gamma + X$ process \cite{Kniehl:2006qq}. Furthermore, the production of $J/\psi$ accompanied by a photon from $Z$ or Higgs boson decays has been systematically investigated \cite{Keung:1983ac,Bodwin:2013gca,Bodwin:2014bpa,Huang:2014cxa,Bodwin:2016edd,Bodwin:2017wdu,Bodwin:2017pzj}. In electron-positron annihilation processes, heavy quarkonium production with an associated photon have also been widely studied \cite{Li:2009ki,Sang:2009jc,Braguta:2010mf,Patel:2011zty,Fan:2012vw,Fan:2012dy,Chao:2013cca,Chen:2013mjb,Li:2013nna,Chen:2013itc,Wang:2013ywc,Sun:2014kva,Xu:2014zra,Chen:2017pyi,Brambilla:2017kgw,Chung:2019ota,Li:2019ncs,Brambilla:2020xod,Yu:2020tri,Sang:2020fql,Liao:2021ifc,Wang:2023ssg} . In 2005, Klasen \textit{et al.} performed the first analytical NLO QCD calculation for $\gamma\gamma \to J/\psi+\gamma$ \cite{Klasen:2004az}, highlighting its promising detectability at the $e^+e^-$ linear collider TESLA. In this article, we extend the theoretical scope by presenting the NRQCD-based NLO calculation of $J/\psi$ polarization in this process at the $e^+e^-$ circular collider CEPC, thereby advancing the understanding of quarkonium-photon associated production.

The paper is structured as follows: Section II details the calculation formalism. Section III presents phenomenological results and analysis. Section IV provides a concise summary.
\section{Calculation Formalism}

Within the NRQCD framework \cite{Bodwin:1994jh,Petrelli:1997ge}, the differential cross section for the process $e^{+}e^{-} \rightarrow e^{+}e^{-}+H(c\bar{c})+\gamma+X$ can be factorized as
\begin{eqnarray}
d\sigma&=&\sum_{n}\int dx_1 dx_2 f_{\gamma}(x_1)f_{\gamma}(x_2)\sum_{i,j}\int dx_i dx_j f_{i/\gamma}(x_i)f_{j/\gamma}(x_j) \nonumber \\
&& \times {d\hat{\sigma}(i+j\to c\bar{c}[n]+\gamma+X)}\times\langle \mathcal O ^{H}(n) \rangle \nonumber
\end{eqnarray}
where $\hat{\sigma}$ denotes the parton-level SDCs for producing the $c\bar{c}$ intermediate state with quantum number $n$ and $\langle \mathcal O ^{H}(n) \rangle$ represents the universal nonperturbative LDMEs. The photon flux $f_{\gamma}(x)$ describes the energy spectrum of bremsstrahlung photons from the initial leptons, while $f_{i/\gamma}(x_i)$ denotes the PDF for parton $i$ (gluon or quark) within a resolved photon. Here, $x$ corresponds to the momentum fraction of the photon relative to the initial electron/positron, and $x_i$ specifies the momentum fraction of parton $i$ relative to the parent photon. At CEPC, the dominant photon source is lepton bremsstrahlung, which can well be described by the Weizsäcker-Williams approximation \cite{Frixione:1993yw} as
\begin{eqnarray}
f_{\gamma}(x)&=&\frac{\alpha}{2\pi}\left[2m^2_e(\frac{1}{Q^2_{max}}-\frac{1}{Q^2_{min}})x + \frac{1+(1-x)^2}{x}\log(\frac{Q^2_{max}}{Q^2_{min}})\right]
\end{eqnarray}
with
\begin{eqnarray}
&&Q^2_{min}=\frac{m^2_e x^2}{1-x}, \nonumber \\
&&Q^2_{max}=\left(\frac{\sqrt{s}\theta}{2}\right)^2(1-x)+Q^2_{min}.
\end{eqnarray}
Here, $x=\frac{E_{\gamma}}{E_e}$ denotes the fraction of the initial electron (or positron) momentum carried by the interacting photon, where 
$\alpha$ is the fine structure constant and $m_e$ the electron mass. The angle $\theta$ between the photon's propagation direction and the lepton beam axis is set to $\theta=32$ mrad as is adopted in the LEP II forward detector. This angular cut defines the phase space for quasi-real photons in the Weizsäcker-Williams approximation.

During the calculations, we set $M_{J/\psi} = 2m_c$ with $m_c = 1.5~\text{GeV}$. The fine structure constant is fixed at $\alpha = \frac{1}{128}$ \cite{Li:2014ava}. As the LO partonic process is a pure QED process, the NLO calculations employ the one-loop $\alpha_s$ running. The renormalization scale is chosen as $\mu_r = \sqrt{p_t^2 + 4m^2_{c}}$. The center-of-mass energy at CEPC is 240 GeV \cite{CEPCStudyGroup:2018rmc,CEPCStudyGroup:2018ghi}. To exclude events where initial-state bremsstrahlung photons directly appear as final-state photons, we impose the criterion $p^{\gamma}_t > p^{\gamma}_{t,\min}$ with $p^{\gamma}_{t,\min} = 1~\text{GeV}$. This transverse momentum cut also ensures the detectability of final-state photons. The CEPC-proposed detector coverage of $|\theta| < 0.99$ \cite{CEPCStudyGroup:2018rmc,CEPCStudyGroup:2018ghi} constrains the photon rapidity to $|y^{\gamma}| < 2.65$.

\begin{table*}[htb]
\caption{The values of the LDMEs.}
\label{LDME}
\begin{tabular}{lccccccc}
\hline\hline
$~~~$ & $\langle \mathcal O ^{J/\psi}(^3S_1^{[1]}) \rangle$ & $\langle \mathcal O ^{J/\psi}(^1S_0^{[8]}) \rangle$ & $\langle \mathcal O ^{J/\psi}(^3S_1^{[8]}) \rangle$ & $\langle \mathcal O ^{J/\psi}(^3P_J^{[8]}) \rangle/m_c^2$\\
$~~~$ & $\textrm{GeV}^3$ & $10^{-2}\textrm{GeV}^3$ & $10^{-2}\textrm{GeV}^3$ & $10^{-2}\textrm{GeV}^3$\\ \hline
set 1 \cite{Butenschoen:2010rq} & $1.32$ & $4.50 \pm 0.72$ & $0.312 \pm 0.93$ & $-0.538 \pm 0.156$\\ \hline
set 2 \cite{Chao:2012iv} & $1.16$ & $8.9 \pm 0.98$ & $0.30 \pm 0.12$ & $0.568 \pm 0.21$\\ \hline
set 3 \cite{Zhang:2014ybe} & $0.645 \pm 0.405$ & $0.78 \pm 0.34$ & $1.0 \pm 0.3$ & $1.78 \pm 0.5$\\ \hline
set 4 \cite{Brambilla:2024iqg} & $1.16 \pm 0.068$ & $0.2489 \pm 0.34$ & $1.050 \pm 0.121$ & $1.879 \pm 0.261$\\ \hline \hline
\end{tabular}
\end{table*}

We employ four representative LDME sets (labeled set~1--4), each providing a consistent description of $J/\psi$ hadroproduction data. Their values, corresponding respectively to the fits in Refs.~\cite{Butenschoen:2010rq, Chao:2012iv, Zhang:2014ybe, Brambilla:2024iqg}, are listed in Table I.

Initially, we analyze the role of resolved photon contributions in the process $\gamma\gamma \to J/\psi + \gamma$. At tree level, we consider the following photon-initiated processes: direct, single-resolved, and double-resolved.
\begin{eqnarray}
\textrm{direct:}&~&\gamma+\gamma \to c\bar{c}[^3S_1^{[1]}]+\gamma; \nonumber \\
\textrm{single~resolved:}&~&\gamma+g \to c\bar{c}[^3S_1^{[8]}]+\gamma; \nonumber \\
\textrm{double~resolved:}&~&g+g \to c\bar{c}[^3S_1^{[1]},^1S_0^{[8]},^3S_1^{[8]},^3P_J^{[8]}]+\gamma, \nonumber \\
&~&q+\bar{q} \to c\bar{c}[^3S_1^{[8]}]+\gamma.
\end{eqnarray}
\begin{figure}
\includegraphics[width=0.45\textwidth]{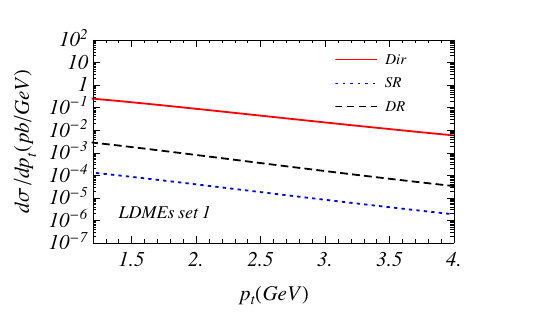}
\includegraphics[width=0.45\textwidth]{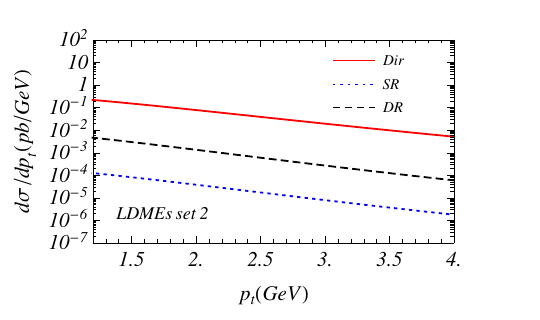}
\includegraphics[width=0.45\textwidth]{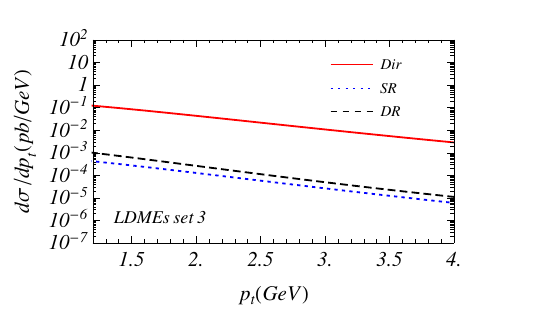}
\includegraphics[width=0.45\textwidth]{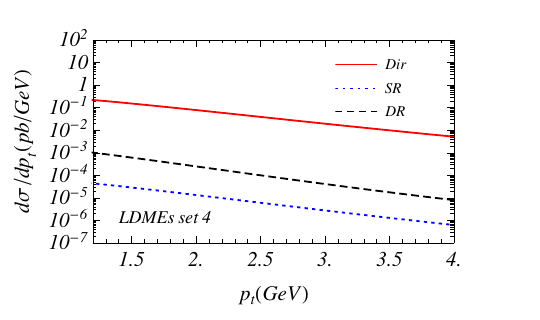}
\caption{\label{fig:resolved}
Differential cross sections for direct ($\mathrm{Dir}$), single-resolved ($\mathrm{SR}$), and double-resolved ($\mathrm{DR}$) processes. For the resolved processes, the GRS99 photon PDF is employed \cite{Gluck:1999ub}.}
\end{figure}
Fig. \ref{fig:resolved} displays the differential cross sections for direct, single-, and double-resolved processes. The direct channel dominates by at least two orders of magnitude, demonstrating that $\gamma\gamma \to J/\psi + \gamma$ proceeds predominantly through direct production. This dominance arises from three key factors: 1) the direct process $\gamma\gamma \to c\bar{c}[^3S_1^{[1]}] + \gamma$ occurs at leading order via color-singlet mechanisms; 2) the single-resolved production $\gamma + g \to c\bar{c}[^3S_1^{[8]}] + \gamma$ suffers strong suppression from $\langle \mathcal{O}^{J/\psi}(^3S_1^{[8]}) \rangle / \langle \mathcal{O}^{J/\psi}(^3S_1^{[1]}) \rangle$; 3) the double-resolved CS process $gg \to c\bar{c}[^3S_1^{[1]}] + \gamma$, mediated by gluon fusion, is suppressed by the small gluon PDF of the photon. In light of these suppression mechanisms, we exclusively analyze the direct channel in subsequent calculations.

Up to $\mathcal{O}(\alpha^3\alpha_s)$ accuracy within the NRQCD factorization, the direct process receives contributions from the Fock states $c\bar{c}[^{3}S_1^{[1]}]$, $c\bar{c}[^{1}S_0^{[8]}]$, and $c\bar{c}[^{3}P_J^{[8]}]$. To be specific,
\begin{eqnarray}
\textrm{LO}~~:~~\gamma+\gamma &\to& c\bar{c}[^3S_1^{[1]}]+\gamma.\nonumber \\
\textrm{NLO}~~:~~\gamma+\gamma &\to& c\bar{c}[^3S_1^{[1]}]+\gamma~(\textrm{virtual corrections}), \nonumber \\
\gamma+\gamma &\to& c\bar{c}[^1S_0^{[8]}]+\gamma+g, \nonumber \\
\gamma+\gamma &\to& c\bar{c}[^3P_J^{[8]}]+\gamma+g. \nonumber
\end{eqnarray}

\begin{figure}
\includegraphics[width=0.65\textwidth]{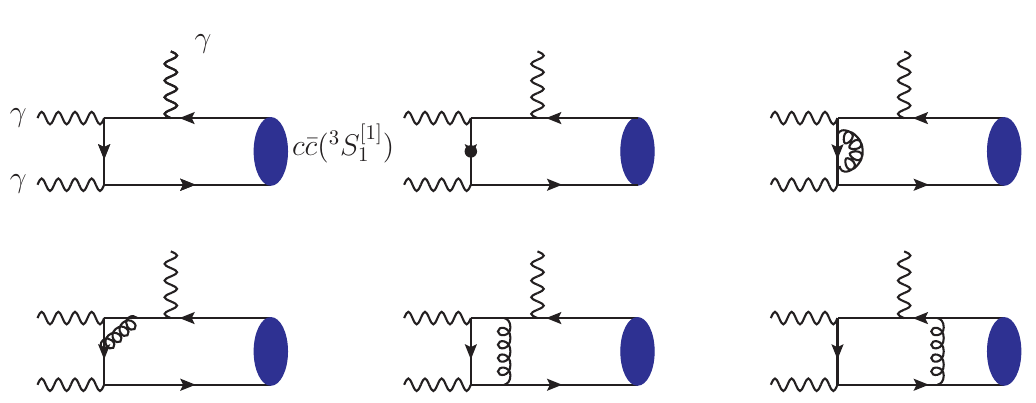}
\caption{\label{fig:Feyn}
Representative Feynman diagrams for $\gamma+\gamma \to c\bar{c}[^3S_1^{[1]}]+\gamma$ production. The second diagram in the first row denotes the counter-term with heavy dot.}
\end{figure}

For the virtual corrections to the process $\gamma + \gamma \to c\bar{c}[^3S_1^{[1]}] + \gamma$, there are 78 Feynman diagrams in total, comprising 48 one-loop diagrams and 30 counter-term diagrams, as shown representatively in Fig. \ref{fig:Feyn}. To isolate the ultraviolet (UV) and infrared (IR) singularities from the one-loop diagrams, dimensional regularization with $D=4-2\epsilon$ is employed. The on-mass-shell (OS) scheme is adopted for the renormalization constants of the quark mass $Z_{\textrm{m}}$ and the quark field $Z_2$, given by
\begin{eqnarray}
\delta Z^{\textrm{OS}}_{\textrm{m}}&=&-3 C_F  \frac{\alpha_s}{4\pi}\left[\frac{1}{\epsilon_{\textrm{UV}}}-\gamma_E+\ln\frac{4\pi \mu_r^2}{m^2}+\frac{4}{3}\right], \nonumber \\
\delta Z^{\textrm{OS}}_2&=&- C_F  \frac{\alpha_s}{4\pi}\left[\frac{1}{\epsilon_{\textrm{UV}}}+\frac{2}{\epsilon_{\textrm{IR}}}-3\gamma_{\textrm{E}}+3\ln\frac{4\pi \mu_r^2}{m^2}+4\right],\nonumber
\end{eqnarray}
where $C_F=\frac{4}{3}$ and $\gamma_{\textrm{E}}$ reprensents the Euler's constant. 

\begin{figure}
\includegraphics[width=0.50\textwidth]{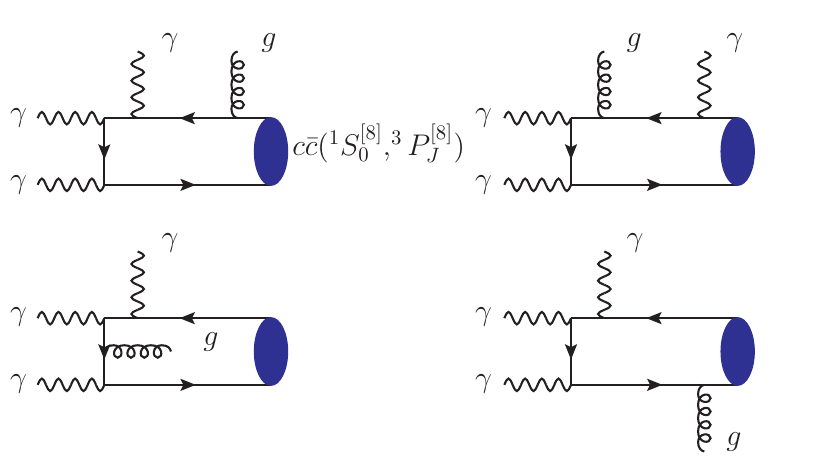}
\caption{\label{fig:Feyn1}
The illustrative diagrams for $\gamma+\gamma \to c\bar{c}[^1S_0^{[8]},^3P_J^{[8]}]+\gamma+g$. }
\end{figure}

For the process $\gamma + \gamma \to c\bar{c}[^1S_0^{[8]}, ^3P_J^{[8]}] + \gamma + g$, there are 24 Feynman diagrams, as shown representatively in Fig. \ref{fig:Feyn1}. The $^1S_0^{[8]}$ channel is divergence free, permitting direct computation. However, in the $^3P_J^{[8]}$ channel, the emitted gluon can be soft, leading to IR singularities. These singularities are absorbed through the renormalization of the corresponding NRQCD matrix elements $\langle \mathcal{O}^{^3P_J^{[8]}}(^3S_1^{[1]}) \rangle$ .\footnote{Owing the absence of $^3S_1^{[8]}$-related processes up to $\mathcal{O}(\alpha^3\alpha_s)$, only $\langle \mathcal{O}^{^3P_J^{[8]}}(^3S_1^{[1]}) \rangle$ requires renormalization.} To derive $\hat{\sigma}_{^3P_{J}^{[8]}}$, we start by decomposing the cross section $\sigma(^3P_{J}^{[8]})$ into two components,
\begin{eqnarray}
\sigma(^3P_{J}^{[8]})=\hat{\sigma}_{^3P_{J}^{[8]}}\langle \mathcal O ^{J/\psi}(^3P_{J}^{[8]})\rangle+\hat{\sigma}^{\textrm{LO}}_{^3S_{1}^{[1]}} \langle \mathcal O ^{^3P_{J}^{[8]}}(^3S_{1}^{[1]})\rangle^{\textrm{renorm}}\langle \mathcal O ^{J/\psi}(^3P_{J}^{[8]})\rangle,
\end{eqnarray}
then one can obtain
\begin{eqnarray}
\hat{\sigma}_{^3P_{J}^{[8]}}\langle \mathcal O ^{J/\psi}(^3P_{J}^{[8]})\rangle
&=&\sigma(^3P_{J}^{[8]})-\hat{\sigma}^{\textrm{LO}}_{^3S_{1}^{[1]}} \langle \mathcal O ^{^3P_{J}^{[8]}}(^3S_{1}^{[1]})\rangle^{\textrm{renorm}} \langle \mathcal O ^{J/\psi}(^3P_{J}^{[8]})\rangle\nonumber \\
&=&\left[\left(\hat{\sigma}_\textrm{S}+\hat{\sigma}_{\textrm{H}}\right)\big|_{^3P_{J}^{[8]}}-\hat{\sigma}^{\textrm{LO}}_{^3S_{1}^{[1]}} \langle \mathcal O ^{^3P_{J}^{[8]}}(^3S_{1}^{[1]})\rangle^{\textrm{renorm}}\right]\langle \mathcal O ^{J/\psi}(^3P_{J}^{[8]})\rangle.
\label{3pj8eq1}
\end{eqnarray}
The small cutoff parameter $\delta_s$ is introduced to separate the phase space into soft ($\hat{\sigma}_{\text{S}}$) and hard ($\hat{\sigma}_{\text{H}}$) regions. The soft part $\hat{\sigma}_{\text{S}}$ can be expressed analytically in terms of the LO SDC $\hat{\sigma}^{\mathrm{LO}}_{{}^3S_1^{[1]}}$ for the partonic process $\gamma\gamma \to c\bar{c}[{}^3S_1^{[1]}] + \gamma$. The hard part $\hat{\sigma}_{\text{H}}$ is infrared-finite and can be evaluated numerically using standard Monte Carlo integration techniques.

The soft singularities in $\hat{\sigma}_\textrm{S}$ cancel against the IR divergences in $\langle \mathcal O ^{^3P_{J}^{[8]}}(^3S_{1}^{[1]})\rangle^{\textrm{renorm}}$, yielding the finite $\hat{\sigma}_{^3P_{J}^{[8]}}$,
\begin{eqnarray}
\hat{\sigma}_{^3P_{J}^{[8]}}=\hat{\sigma}_{\textrm{H}}-\frac{4\alpha_s}{3 \pi m^2_c}u_{\epsilon}\hat{\sigma}^{\textrm{LO}}_{^3S_{1}^{[1]}},
\label{3pj8 SDCs}
\end{eqnarray}
with \cite{Zhang:2014coi}
\begin{eqnarray}
u_{\epsilon}=\frac{p_0}{|\bf{p}|}\textrm{ln}\left(\frac{p_0+|\bf{p}|}{p_0-|\bf{p}|}\right)+\textrm{ln}\left(\frac{\mu_{\Lambda}^2}{s\delta_s^2}\right)-2+2\textrm{ln}(2), 
\end{eqnarray}
where $p_0$ and $\mathbf{p}$ denote the energy and three-momentum of the $J/\psi$, respectively. $\mu_\Lambda$ denotes the scale arising from the renormalization of the LDME and is set to $\mu_\Lambda = m_c$.

In our calculations, we employ our Mathematica-Fortran package, which implements FeynArts \cite{Hahn:2000kx}, FeynCalc \cite{Mertig:1990an}, FIRE \cite{Smirnov:2008iw}, and Apart \cite{Feng:2012iq}. This package has been used for several heavy-quarkonium processes in $\gamma\gamma$ and $\gamma p$ collisions \cite{Sun:2017nly,Sun:2017wxk,Sun:2015hhv}. Additionally, we utilize the independent Feynman Diagram Calculation (FDC) package \cite{Wang:2004du} to compute all relevant processes, obtaining consistent numerical results. By coherently superimposing the bremsstrahlung \cite{Frixione:1993yw} and beamstrahlung \cite{Chen:1993dba} photon spectra as described in Ref.~\cite{Klasen:2004az}, we have performed a comparison with the results therein. We observe a discrepancy: our LO and NLO cross sections are systematically lower by $15\%$--$35\%$ compared to Fig.~8(a) of Ref.~\cite{Klasen:2004az}. Nevertheless, the $K$-factors we obtain show trends similar to those in its Fig.~9(a). As a crosscheck of our computational framework, we reproduced the LO and NLO cross sections for the direct process $\gamma\gamma \to c\bar{c}[^3S_1^{[1]}]+\gamma$ from Fig.~3 of Ref.~\cite{Yedelkina:2023cbu} under identical kinematic cuts.

\section{Phenomenological results}

In this section, we present our numerical results of the process $\gamma\gamma \to J/\psi+\gamma$, which exclusively consider the direct photon production at NLO accuracy and exclude the resolved photon contributions.

\begin{table*}[htb]
\caption{Integrated cross sections (in unit: pb) of $J/\psi$ in $\gamma\gamma \to J/\psi+\gamma+X$ under the kinematic cuts $p_t^{\gamma} > 1~\textrm{GeV}$ and $|y^{\gamma}| < 2.65$, with the charm quark mass fixed at $m_c = 1.5~\textrm{GeV}$.}
\label{int sigma}
\begin{tabular}{lccccccc}
\hline\hline
$~~~$ & $\textrm{CS}_{\textrm{LO}}$ & $\textrm{CS}_{\textrm{NLO}}$ & NRQCD\\ \hline
set 1~~~ & $0.224$ & $0.108$ & $0.112$\\ \hline
set 2~~~ & $0.197$ & $0.095$ & $0.096$\\ \hline
set 3~~~ & $0.110$ & $0.053$ & $0.049$\\ \hline
set 4~~~ & $0.197$ & $0.095$ & $0.091$\\ \hline\hline
\end{tabular}
\end{table*}

The integrated cross sections of $J/\psi+\gamma$ production in $\gamma\gamma$ collisions, computed for four LDME sets, are presented in Table \ref{int sigma}. ``CS" denotes the contribution of $c\bar{c}[^3S_1^{[1]}]$, while ``NRQCD" represents the complete calculation including both CS and CO ($^1S_0^{[8]}$, $^3P_J^{[8]}$) channels. The tabulated data reveal that NLO corrections to the $^3S_1^{[1]}$ process significantly suppress the LO results, with a reduction magnitude approaching 50\%. Contributions from CO states exhibit slight impact on the integrated cross section. Assuming the nominal integrated luminosity of 300~fb$^{-1}$ \cite{CEPCStudyGroup:2018rmc,CEPCStudyGroup:2018ghi} and considering the dileptonic branching fraction $\mathcal{B}(J/\psi \to \ell^+\ell^-) \approx 10\%$, CEPC is expected to accumulate $\mathcal{O}(10^3)$ observable $J/\psi$ events per year.

\begin{figure}
\includegraphics[width=0.45\textwidth]{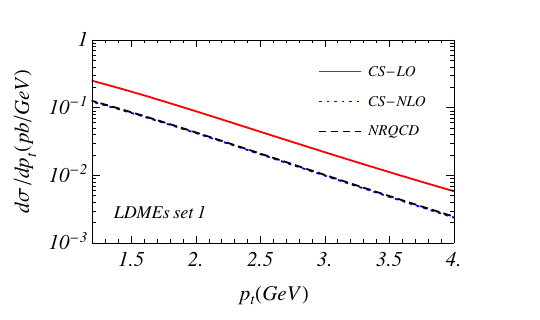}
\includegraphics[width=0.45\textwidth]{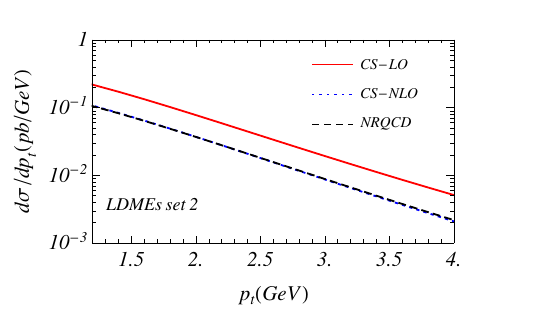}
\includegraphics[width=0.45\textwidth]{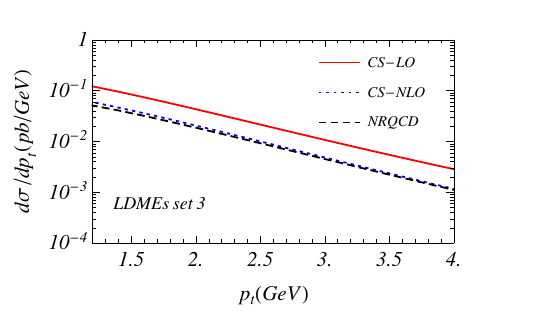}
\includegraphics[width=0.45\textwidth]{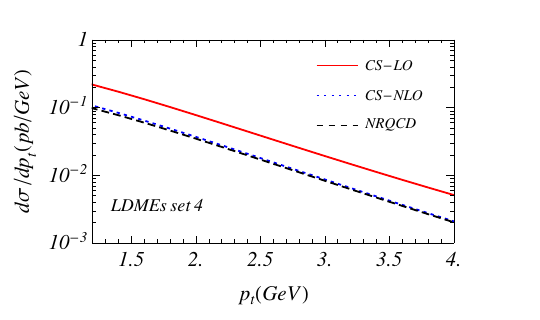}
\caption{\label{fig:pt dis}
The $p_t$ distributions for $J/\psi + \gamma$ production in $\gamma\gamma$ collisions.}
\end{figure}

The $p_t$ distributions in Fig. \ref{fig:pt dis} compare NRQCD predictions with CS results, demonstrating a rapid decrease in differential cross section $d\sigma/dp_t$ with increasing transverse momentum. Slight CO contributions cause near-overlap between CS and full NRQCD curves.

Subsequently, we investigate the polarization distribution of $J/\psi$ in the process $\gamma\gamma \to J/\psi + \gamma + X$. This work focuses exclusively on the polarization parameter $\lambda_\theta$ in the helicity frame, defined as \cite{Beneke:1998re}:
\begin{eqnarray}
\lambda_\theta=\frac{d\sigma_{11}-d\sigma_{00}}{d\sigma_{11}+d\sigma_{00}},
\end{eqnarray}
where $d\sigma_{S_z,S^{'}_z}(S_z,S^{'}_z=0,\pm1)$ denote the spin density matrix elements.

\begin{figure}
\includegraphics[width=0.45\textwidth]{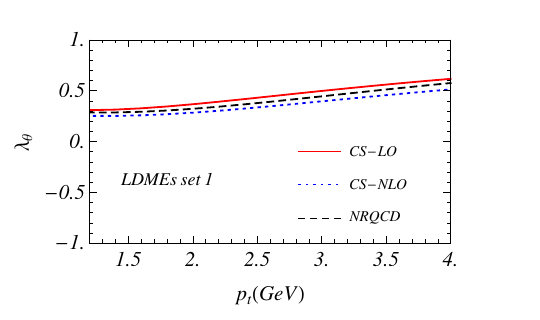}
\includegraphics[width=0.45\textwidth]{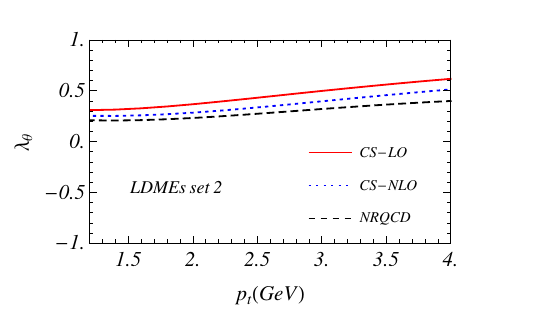}
\includegraphics[width=0.45\textwidth]{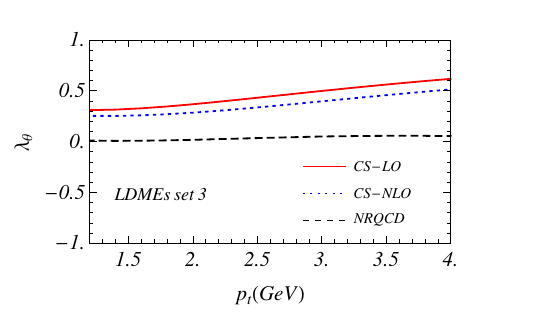}
\includegraphics[width=0.45\textwidth]{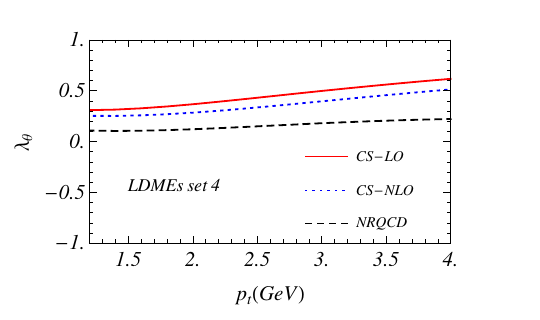}
\caption{\label{fig:pol}
Comparison of $J/\psi$ polarization predictions between the CS model and full NRQCD factorization.}
\end{figure}

Fig. \ref{fig:pol} reveals close agreement between CSM and full NRQCD polarization predictions for LDME sets 1 and 2. By contrast, a significant discrepancy emerges for sets 3 and 4: while the CSM yields pronounced transverse polarization, the full NRQCD calculation gives nearly unpolarized production for set 3 and only weakly transverse polarization for set 4. In consideration of the consistent yield predictions between CS and NRQCD frameworks, the polarization divergence in sets 3 and 4 necessitates investigation into its physical origin.

\begin{figure}
\includegraphics[width=0.45\textwidth]{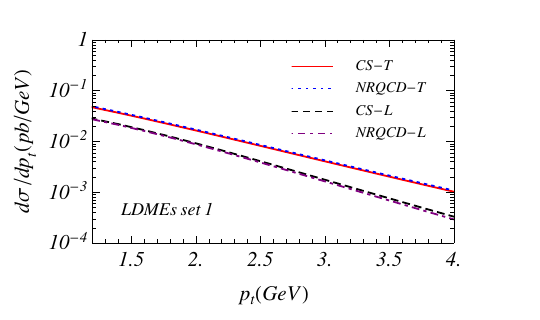}
\includegraphics[width=0.45\textwidth]{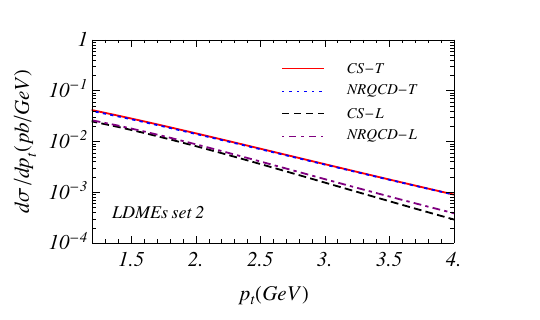}
\includegraphics[width=0.45\textwidth]{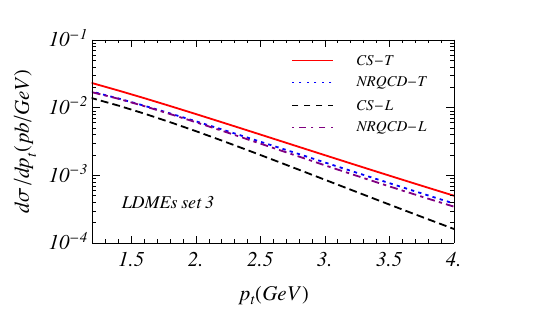}
\includegraphics[width=0.45\textwidth]{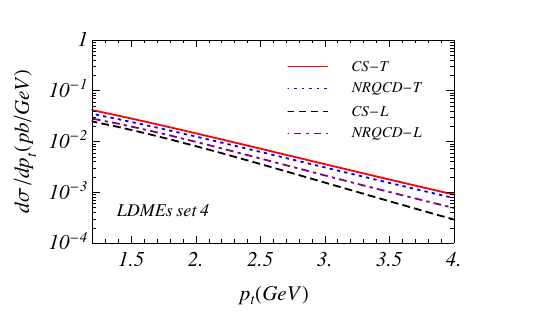}
\caption{\label{fig:pol_LT}
The $p_t$ distributions for $J/\psi + \gamma$ production in $\gamma\gamma$ collisions. ``T'' denotes transverse-polarized cross sections, ``L'' longitudinal-polarized components.}
\end{figure}

To address this, we decompose the differential cross sections into transverse (T) and longitudinal (L) components, as compared in Fig. \ref{fig:pol_LT}. For sets 1 and 2, CO contributions marginally perturb both the transverse and longitudinal CS components. For sets 3 and 4, however, CO terms substantially suppress the ``T" parts while enhancing ``L" parts, thereby transforming the initially transverse-polarized state into near unpolarization. Equation \ref{3pj8 SDCs} reveals that $\hat{\sigma}_{^3P_J^{[8]}}$ contains a negative term proportional to $\hat{\sigma}_{^3S_1^{[1]}}^{\textrm{LO}}$. Owing to the transverse dominance in $\hat{\sigma}_{^3S_1^{[1]}}^{\textrm{LO}}$, $\hat{\sigma}_{^3P_J^{[8]}}$ therefore develops negative transverse and positive longitudinal components. For LDME sets 1 and 2, the small ratio $R=\frac{\langle \mathcal O^{J/\psi}(^3P_J^{[8]})\rangle}{\langle \mathcal O^{J/\psi}(^3S_1^{[1]})\rangle}$ leads to modest polarization modifications. In contrast, sets 3 and 4 exhibit a relatively large ratio $R$, where the $^3P_J^{[8]}$ contribution significantly suppresses the transverse component while enhancing the longitudinal one, subsequently drastically altering the CS polarization pattern. Crucially, since the $^1S_0^{[8]}$ contribution is negligible and unpolarized, and $^3S_1^{[8]}$ states are absent at current $\alpha_s$ order, $J/\psi$ polarization in $\gamma\gamma \to J/\psi+\gamma+X$ is solely sensitive to $R$. The persistent challenges in describing $J/\psi$ polarization largely stem from the difficulties in determining the LDMEs. A key advantage of the process $\gamma\gamma \to J/\psi + \gamma + X$ over hadroproduction is its selective sensitivity: while the latter receives essential contributions from all three CO states ($^1S_0^{[8]}$, $^3S_1^{[8]}$, and $^3P_J^{[8]}$), polarization in the former depends almost solely on the $^3P_J^{[8]}$ LDME, consequently providing a cleaner probe for constraining this specific matrix element and improving the precision of NRQCD predictions.

\begin{figure}
\includegraphics[width=0.65\textwidth]{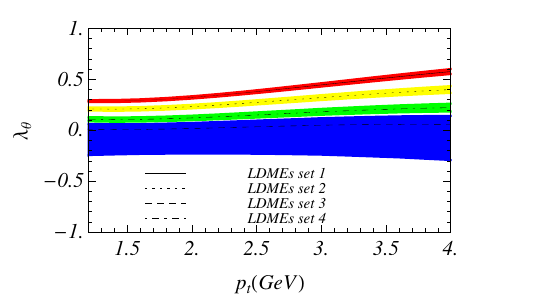}
\caption{\label{fig:pol_bands}
$J/\psi$ polarization predictions. The bands are given by the LDMEs uncertainties.}
\end{figure}

Finally, Fig. \ref{fig:pol_bands} compares polarization parameters predicted by the four LDME sets, incorporating their respective uncertainties. The uncertainty bands reveal distinct polarization patterns:
\begin{itemize}
\item
sets 1 and 2 predict moderately to strongly transverse polarization.
\item
set 3 yields near-unpolarized or weakly longitudinal behavior.
\item
set 4 produces near-unpolarized to slightly transverse polarization.
\end{itemize}
Seeing these significant discrepancies, $\gamma\gamma \to J/\psi + \gamma+X$ production serves as a critical benchmark process to test the validity of LDMEs extracted from hadroproduction data.
\section{Summary}
Within the NRQCD factorization framework, we systematically investigate $\gamma\gamma \to J/\psi + \gamma$ production. Our analysis employs four representative LDME sets that consistently describe $J/\psi$ hadroproduction data, with QCD corrections applied specifically to the direct-photon channel. The results demonstrate absolute dominance of the direct process, exceeding single- and double-resolved contributions by over two orders of magnitude; at CEPC's design luminosity, this yields $\mathcal{O}(10^3)$ reconstructed $J/\psi$ events annually. NLO corrections reduce the CS cross section by $\sim$50\% while CO states show moderate contributions. For polarization, the CS model predicts pronounced transverse polarization. In sharp contrast, the NRQCD predictions exhibit strong LDME dependence: while two sets yield results consistent with the CSM, another set generates near-unpolarized to moderately longitudinal behavior, and the remaining one produces near-unpolarized to slightly transverse polarization. This spread highlights a significant theoretical divergence. Furthermore, the polarization predictions depend critically on the ratio $\frac{\langle \mathcal{O}^{J/\psi}(^3P_J^{[8]})\rangle}{\langle \mathcal{O}^{J/\psi}(^3S_1^{[1]})\rangle}$, but not on the specific values of $\langle \mathcal{O}^{J/\psi}(^1S_0^{[8]})\rangle$ and $\langle \mathcal{O}^{J/\psi}(^3S_1^{[8]})\rangle$. Thus, $\gamma\gamma \to J/\psi + \gamma + X$ provides a critical probe for testing LDME universality and resolving the longstanding $J/\psi$ polarization puzzle.
\section{Acknowledgments}

\noindent{\bf Acknowledgments}:
This work is supported in part by the Natural Science Foundation of China under the Grant No. 12065006, the Guizhou Provincial Top-Quality Course under the Grant No. 2024JKXX0048 and the Projectof Guizhou Provincial Department of Science and Tech-nology under Grant No.CXTD[2025]030.\\

\end{document}